\documentclass[aps,pre,twocolumn,superscriptaddress,showpacs]{revtex4}

\usepackage{amsmath}
\usepackage{amsfonts}
\usepackage{graphicx}

\graphicspath{{./pictures/}}

\begin{document}


\title{Convex Hulls of Multiple Random Walks: A Large-Deviation Study}


\author{Timo Dewenter}
\affiliation {Institut f\"ur Physik, Universit\"at Oldenburg, D-26111
Oldenburg, Germany}

\author{Gunnar Claussen}
\affiliation {Institut f\"ur Physik, Universit\"at Oldenburg, D-26111
Oldenburg, Germany}
\affiliation {Fachbereich Ingenieurwissenschaften, Jade Hochschule 
Wilhelmshaven/Oldenburg/Elsfleth, D-26389 Wilhelmshaven, Germany}

\author{Alexander K. Hartmann}
\email[]{a.hartmann@uni-oldenburg.de}
\homepage[]{http://www.compphys.uni-oldenburg.de/en/}
\affiliation {Institut f\"ur Physik, Universit\"at Oldenburg, D-26111
Oldenburg, Germany}

\author{Satya N. Majumdar}
\email[]{satya.majumdar@u-psud.fr}
\affiliation{Laboratoire de Physique Th\'{e}orique et 
Mod\`{e}les Statistiques (UMR 8626 du CNRS), 
Universit\'{e} de Paris-Sud, B\^{a}timent 100, 
91405 Orsay Cedex, France}

\date{\today}

\begin{abstract}
We study the polygons governing the convex hull of a point set created by 
the steps of $n$ independent two-dimensional random walkers. Each such 
walk consists of $T$ discrete time steps, where $x$ and $y$ increments are 
i.i.d.~Gaussian. We analyze  area $A$ and  
perimeter $L$ of the convex hulls. We obtain probability densities for these two quantities 
over a large range of the support by using a large-deviation approach 
allowing us to study densities below $10^{-900}$. We find that the densities 
exhibit a universal scaling behavior as a function of 
$A/T$ and $L/\sqrt{T}$, respectively.
As in the case of one walker ($n=1$), the 
densities follow Gaussian distributions for $L$ and $\sqrt{A}$, 
respectively. We also obtained the rate functions for the area and 
perimeter, rescaled with the scaling behavior of their maximum
possible values, and found limiting 
functions for $T \rightarrow \infty$, revealing that the densities follow 
the large-deviation principle. These rate functions can be described by 
a power law for $n \rightarrow \infty$ as found in the 
$n=1$ case. We also investigated 
the behavior of the averages as a function of the number of walks $n$ and 
found good agreement with the predicted behavior.
\end{abstract}


\maketitle

\section{Introduction \label{intro}}
Originally, random walks have been introduced by P\'olya \cite{Polya1921} 
in 1921. Since then, many studies have dealt with this topic, as they 
are an ubiquitous model for physical, biological and social
processes \cite{vanKampen1992,bergHC1993,hughes1996}. 
Example applications from biology include self-propelled motion of 
bacteria, and the diffusion of nutrients \cite{bergHC1993}, as well as 
animal motion in general \cite{Bovet1988,Bartumeus2005}. Another example  
is the marking of territories by animals or the description of home ranges 
\cite{mohr1947,worton1995,Giuggioli2011}. 
For the latter case a strong increase of the amount of experimentally available 
data ocurred after the introduction of
automated radio/GPS tagging of animals \cite{kenward1987,landguth2010}.
The usage of minimum convex polygons, called \emph{convex hulls}, bordering the
trace of an animal \cite{mohr1947,worton1987} 
is a simple yet versatile \cite{boyle2009} way to describe
the home range and can be used for any type of (random-walk) data. 
In two dimensions, the convex hull of a point 
set is the minimum subset whose elements form a convex polygon in 
such a way that (a) all points of the set and (b) the connecting 
lines between all possible pairs lie inside the polygon.

Much progress has been made on the analytical side, when the number of 
steps is very large and the random walk (with a finite variance of the 
step size) converges to the continuous-time Brownian motion (for a review
see e.g., \cite{Majumdar2010}). The mean perimeter 
\cite{Takacs1980,Letac1993} and mean area \cite{ElBachir1983}
of a single two-dimensional Brownian motion are known for a long time.

It was shown \cite{Randon-Furling2009,Majumdar2010} recently that the 
problem of computing 
the mean perimeter and the mean area of the convex hull of an arbitrary 
two-dimensional stochastic process can be mapped to computing the extremal 
statistics of the one-dimensional component of the process. This procedure 
was successfully applied recently to compute the mean perimeter and the 
mean area of several two-dimensional stochastic processes such as the 
random acceleration process in 2D~\cite{Reymbaut2011}, 2D branching 
Brownian motions with absorption and applications to edpidemic 
outbreak~\cite{Dumonteil2013} and 2D anomalous diffusion 
processes~\cite{Lukovic2013}.
Very recently, this method was also successfully used to compute the 
exact mean perimeter of the convex hull of a planar Brownian motion 
confined to a half-space~\cite{Chupeau2014}.
Finally, using different methods, the mean perimeter and the mean area of 
the convex hull of a single Brownian motion, but in arbitrary dimensions, 
have been computed recently in the mathematics 
literature~\cite{Eldan2011,Kabluchko2014}.

Analytical calculations of even the second moment for the area and 
perimeter of a convex hull regarding single two-dimensional Brownian 
motion turned 
out to be very difficult \cite{Snyder1993,Goldman1996}. For the full 
distributions of the area and perimeter no analytical results are known 
so far, so the usage of computer simulations is a natural approach, as 
done in a recent study \cite{Claussen2015}.

Here, we are interested in multiple random walkers, which perform their
walks independently from each other. The investigation of $n$ 
non-interacting random 
walkers on a $d$-dimensional regular lattice has been done in 
\cite{Larralde1992,Larralde1992a,Yuste1999}. 
Many studies have been published for 
interacting multiparticle walkers \cite{Acedo2002}, e.g., in one 
dimension \cite{Forrester1989,Kukla1996,Aslangul1999,Schehr2008,
Forrester2011,Rambeau2011,Schehr2013,Kundu2014}. The mean 
first passage time of $n$ independent diffusing particles in Euclidean 
space is calculated, e.g., in \cite{Draeger1999} or \cite{Krapivsky2010}. 
Recently, the mean 
perimeter and the mean area of $n$ independent Brownian motions have been 
computed in 2D~\cite{Randon-Furling2009,Majumdar2010}. In this article we 
want to check the predictions from theory by numerical simulations of $n$
non-interacting time-discrete Brownian random walkers as well as the 
probability density functions of the area and perimeter of the 
corresponding convex hulls. 
In particular we apply a numerical large-deviation approach
to obtain the probability density functions over a large range of the
support, down to probability densities as small as $10^{-900}$.
In addition, we are interested in what way 
the same results for $n>1$ walkers are found in comparison to the $n=1$ 
case \cite{Claussen2015}.

The paper is organized as follows: Section \ref{RW_CH_A} introduces the 
random walk model, the convex hull of a two-dimensional point set, and 
briefly elucidates an algorithm to obtain such a convex hull. Part 
\ref{LD_sch} explains the large-deviation scheme used to obtain the 
probability density function over a large range of the support including 
the low-probability tails. The next part \ref{res} presents the results 
achieved from our simulations. The last section \ref{Concl} concludes the 
article and a short outlook is given.

\section{Random Walks, Convex Hulls, and Algorithms \label{RW_CH_A}}
A time-discretized random walk consists of $T$ step vectors 
$\vec{\delta}_i$, and the position $\vec{x}(\tau)$ at time step 
$\tau < T$ is the sum of all steps up to $\tau$, i.e.:
\begin{equation}
 \vec{x}(\tau) = \vec{x}_0 + \sum^{\tau}_{i=1}\vec{\delta}_i.
 \label{pos}
\end{equation}
The walk configuration itself is then the set 
$\mathcal{W} = \{\vec{\delta}_1, \vec{\delta}_2, ..., \vec{\delta}_T\}$ 
of steps \cite{Feller1950}. The step $\vec{\delta}_i = 
(\delta_{x,i}, \delta_{y,i})$ itself denotes a displacement of the 
particle by $\delta_{x,i}$ in $x$-direction and $\delta_{y,i}$
in $y$-direction. Here, we consider a time-discrete approximation to a 
Brownian walk, i.e., both 
$\delta_{x,i}$ and $\delta_{y,i}$ are, for each $i$, drawn randomly 
from a Gaussian distribution with zero mean and variance one.
All 
considered walks are open, i.e., the walker does not need to get back 
to the starting point $\vec{x}(0)$ after $T$ steps.

In contrast to \cite{Claussen2015}, where only single walks with one 
walker have been 
investigated, we put multiple random walks under scrutiny. So, starting 
from the origin of the coordinate system, $n$ independent random walkers 
perform their walks simultaneously. The resulting point set 
$\mathcal{\widetilde W}$
of $n \cdot T$ points given by the 
individual positions of all $n$ walkers after each time step is then 
further investigated.

The convex hull $\mathcal{C} = \text{conv}(\mathcal{\tilde{P}})$ of a 
two-dimensional point set $\mathcal{\tilde{P}} = \{ \tilde{P_i} \}, 
\tilde{P_i} \in \mathbb{R}^2$ 
is described through a convex set over $\mathcal{\tilde P}$. The points $P$ 
within $\mathcal{C}$ are given by all possible combinations 
$P = \sum_i \alpha_i \tilde{P_i}$ with $\tilde{P_i} \in \mathcal{\tilde P}$ and 
$\sum_i \alpha_i = 1$ and $\alpha_i \in \mathbb{R}_0^+$ (definition 
given according to \cite{Preparata1985}). This means:
\begin{enumerate}
 \item All points $\tilde{P}_i \in \mathcal{\tilde{P}}$ lie within $\mathcal{C}$.
 \item All lines $\overline{\tilde{P}_i \tilde{P}_j}; \tilde{P}_i, \tilde{P}_j 
 \in \mathcal{\tilde{P}}$ also 
lie within $\mathcal{C}$.
\end{enumerate}

The boundary of the convex set is a 
polygon which connects a subset $\mathcal{P}\subset \mathcal{\tilde P}$
of $H$ points from the point set, i.e.,  
$\mathcal{P}=\{P_0,P_1,\ldots,P_{H-1}\}$, with $P_i=(x_i,y_i)$ 
($i=0,\ldots,H-1$). The hull is attributed with area 
$A$ and perimeter $L$ according to (identifying $i=H$ with $i=0$):
\begin{align}
 A(\mathcal{C}) = \frac{1}{2} \sum_{i=0}^{H-1} (y_i + y_{i+1}) (x_i-x_{i+1}) \\
 L(\mathcal{C}) = \sum_{i=0}^{H-1} \sqrt{(x_i-x_{i+1})^2 + (y_i-y_{i+1})^2}
\end{align}

For our work, we determined the polygons bordering convex hulls
(for which one uses shortly the  term ``convex hull'') numerically. 
For convenience, we use dimensionless quantities subsequently, as all 
convex hulls are represented in a computer.

Here, we used the ``Jarvis March'' algorithm \cite{Jarvis1973}, 
which has a complexity of $\mathcal{O} (N \cdot H)$, where $N$ is the 
number of points in the investigated point set and $H$ the number of 
points in the convex hull. In this algorithm, the convex hull is 
calculated in a ``gift-wrapping'' manner, where one needs to make sure 
that all points of the set lie on e.g., the right side of a starting 
point. The next point added to the convex hull is the point which has 
the minimum angle between the line connecting both points and the 
vertical. This procedure is repeated until one reaches the starting 
point again.

In usual cases, the application of convex hull algorithms can be 
accelerated by usage of pre-selection heuristics, such as the 
one introduced by Akl and Toussaint \cite{Akl1978}. This heuristic looks 
up extreme points of the set (i.e., those of maximum and minimum $x$- 
and $y$-coordinates) and discards all points which lie inside 
the quadrilateral formed by these points. We use a custom refinement 
of this heuristic, which is based on iterating the heuristic under 
rotation of the coordinate origin, which eliminates another fraction 
of inert points per each iteration.

\section{Large-Deviation Scheme \label{LD_sch}}
For simple-sampling results, walk configurations $\mathcal{W}$ for 
multiple walkers $n$
are generated randomly, and the according convex hulls $\mathcal{C}$ 
are calculated through the algorithm, resulting in a multitude 
of values of 
$A$ and $L$. Obtaining histograms of these
values only gives access to the high probability regime, where
the convex-hull properties of typical random walks are measured.
However, in order to obtain values of these quantities 
with especially low probabilities, allowing us to measure the distributions
$P_n(A)$ and $P_n(L)$ over a large range of the support, 
a certain Markov-Chain Monte 
Carlo (MCMC) scheme can be used \cite{align2002,Hartmann2011}.

The MCMC consists of an evolution of random walks $\mathcal{W}(t)$
and corresponding  sets $\mathcal{\widetilde W}(t)$ of points.
$t$ is another discrete ``time'' parameter, not to be confused
with the time parameter $\tau$ of the random walks.
For the walks, we measure the property $S(t)$, i.e., the area ($S=A$) or 
perimeter ($S=L$) of the convex hull of the point sets, 
depending on which distribution $P_n(A)$ or $P_n(L)$ we are aiming at.
The initial configuration $\mathcal{W}(0)$ is any walk configuration,
e.g., a randomly chosen one.

At each Monte Carlo step $t$, all $n$ independent walks 
$\mathcal{W}_k(t)$ ($k \in \{1,2,\ldots,n\}$) are altered to 
$\mathcal{W}_k^*$ by replacing one randomly selected step $\vec{\delta}_i$
($i\in\{1,2,\ldots,T\}$) in each walk 
with a newly generated step $\vec{\delta}'_i$. The new step is generated
according to the same distribution as all other random walk steps, i.e., 
the $x$- and $y$-coordinate of $\vec{\delta}'_i$ are drawn independently 
from a Gaussian distribution. Note that by exchanging e.g., the first step 
$\vec{\delta}_1$, all following positions $\vec{x}$ 
(cf., Eq.\ \eqref{pos}) of the walk are changed. 
The convex hull of the point set 
$\mathcal{\widetilde W}^* = \bigcup_{k} \mathcal{\widetilde W}_k^*$ 
resulting from the $n$ walks
$\mathcal{W}_k^*$ is calculated, leading to the quantity $S^*$. 
The alteration $\mathcal{W}^*$ is \emph{accepted} 
($\mathcal{W}(t+1) = \mathcal{W}^*$) 
according to the Metropolis probability:

\begin{equation}
 p_{\text{Met}} = \min\left[1, e^{-(S^*-S(t))/\Theta} \right]\,.
\end{equation}
Here, $\Theta$ is the (artificial) Monte Carlo ``temperature'', which is
a parameter used to set the range of the sampled values. 
If the alteration is not accepted, it is \emph{rejected}, i.e., 
$\mathcal{W}(t+1) = \mathcal{W}(t)$.

Like in any MCMC simulation one needs to equilibrate the simulation, i.e.,
discards the initial part of the measured quantities until ``typical''
values are found. Typical equilibration times are $10^3$ sweeps (one sweep 
equals $T$ MC steps) for e.g., $T=200$, $n=3$, and $\Theta =10$ for the area of 
the convex hull. In addition, we pick only each $k$th data point from the 
original measurement to get roughly decorrelated values. 
For the case above we use $k = 1$ sweeps, which is a 
typical value. Towards low absolute values of the temperature this value 
needs to be increased, so e.g., for $\Theta= 0.2$ in the above case, we choose 
$k=100$ sweeps.

For a given quantity ($S=A$ or $S=L$) and a given walk length $T$ one gets 
different probability density functions (pdfs) $P_{\Theta}(S)$ for each 
temperature $\Theta$ used. They are related to the 
actual distribution $P(S)$ according to the relation \cite{align2002}
\begin{equation}
 P(S) = e^{S/\Theta} \ Z(\Theta) \ P_{\Theta}(S)\,,
\end{equation}
where $Z(\Theta)$ is a normalization constant.
For different values of $\Theta$, different ranges of the
measured value $S$ are obtained. This allows for a piecewise
reconstruction of $P(S)$ via suitable choices of the
normalization constants 
$Z(\Theta)$. They can be calculated through inversion of this 
formula whenever for two values $\Theta_1$ and $\Theta_2$ the
ranges of the sampled values of $S$ overlap. Thus, the temperatures
are chosen such that for neighbouring $\Theta$ the measured
histograms sufficiently overlap.

For a more detailed description of the calculation of the 
normalization constants $Z(\Theta)$ and the determination of the pdf from 
the pdfs for the single temperatures we refer to, e.g., 
\cite{align2002,Hartmann2004}.

Note that the large-deviation approach has already been applied 
successfully for the
case of the convex hull of the point set of 
one ($n=1$) walker \cite{Claussen2015}. In that reference also the test case
of independent points was simulated and a comparison with analytical results
yielded a good agreement.

\section{Results \label{res}}
For $n=3$ random walks we performed simulations for walk lengths of 
$T \in [20, 200]$ while measuring and biasing 
for the area $A$ and the perimeter $L$ of the convex 
hulls, respectively.
To obtain a large range of the support for the pdfs of these two 
quantities we used e.g.\ 17 temperatures $\Theta \in [-200,40]$ (excluding the 
value $\Theta = \infty$, which corresponds to simple sampling) for $T=200$ for 
the pdf of $A$ and about 40 temperatures $\Theta \in [-20,5]$ for $L$.

We also studied the case of $n=2$ walks, which is closer
to the single walker case. Here, we used walks of lengths $T \in [20,500]$. 

To investigate 
the behavior with increasing number of walks $n$, due to the strongly
increasing numerical effort, we performed  
simulations at fixed system size $T=50$ and variable number of walks 
$n \in [2,6]$ for both observables $A$ and $L$. We again obtained 
probability density functions over a large range of the support. 
In addition, we performed 
simple-sample simulations, i.e., close to the peak of the histogram, for 
$T=50$, $10^6$ samples and up to $n=10^5$ independent random walkers.

\begin{figure}[htb]
 \includegraphics[width=0.45\textwidth]{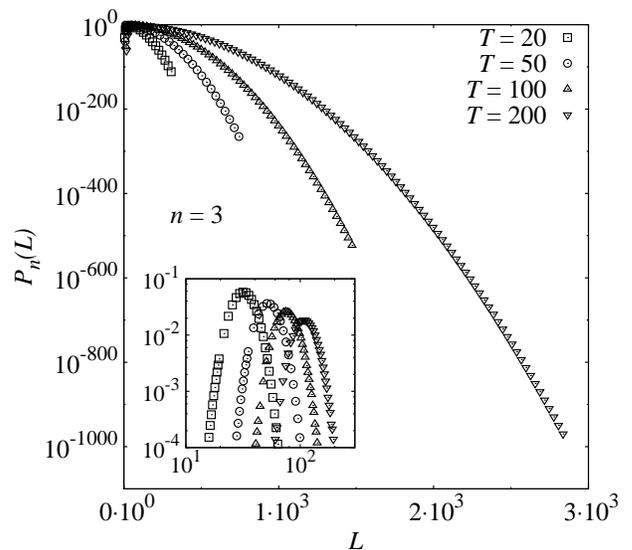}
\caption{Probability density function $P_n(L)$ of the perimeter~$L$ of the 
convex hull of $n=3$ independent random walks in semi-logarithmic scale. 
Inset: Region around peaks in double logarithmic scale. 
\label{pdf_3walks_peri}}
\end{figure} 

\subsection{Probability density function}
\label{pdf}
As an example, Fig.\ \ref{pdf_3walks_peri} shows the pdf of the perimeter of 
the convex hull of three independent two-dimensional time-discrete open 
Brownian walks. By using the large-deviation approach, probability 
densities smaller than $10^{-900}$ can be reached. One can observe the 
strong curvature of the data on a semi-logarithmic scale. With increasing 
walk length $T$ the probability densities also increase when looking at a 
fixed perimeter. This is due to the fact that for larger walk lengths 
large perimeters are found by the simulations more likely as more steps 
in the random walk are available. We obtained results with similar 
high numerical quality for the probability density of the area 
(not shown without rescaling for brevity).


\begin{figure}[ht]
 \includegraphics[width=0.45\textwidth]{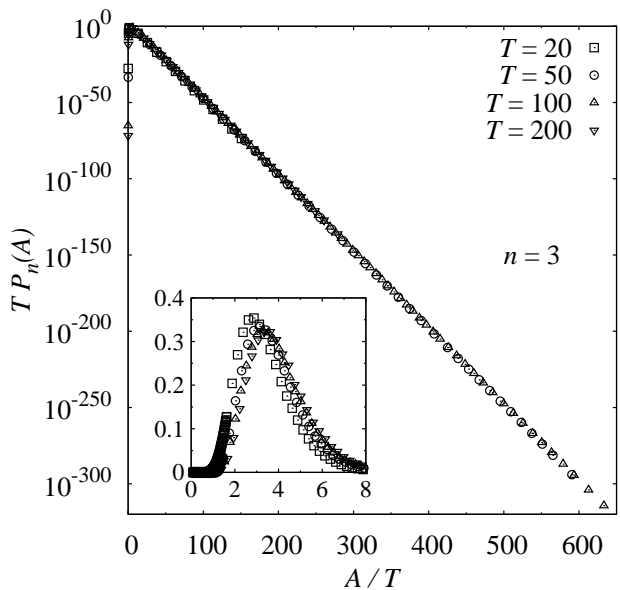}
\caption{Rescaled pdfs for $n=3$ walks, the area $A$ of the convex hull 
 and different walk lengths $T$ in semi-logarithmic scale. Inset: Region 
 close to peaks in linear scale. \label{pdf_3walks_area_res}}
\end{figure}

Next, we check whether the scaling assumptions for the area \cite{Majumdar2010}
\begin{equation}
 P_{T,n}(A) = \frac 1 T \; \widetilde{P}_n \left( \frac A T \right),
 \label{scaling_A}
\end{equation}
and the perimeter 
\cite{Majumdar2010}
\begin{equation}
  P_{T,n}(L) = \frac{1}{\sqrt T} \; \widetilde{P}_n \left( \frac{L}{\sqrt T} \right),
  \label{scaling_M}
\end{equation}
are also valid \cite{Claussen2015} in the case of multiple ($n=2,3$) random 
walks. Here $\widetilde{P}_n (\cdotp)$ are universal 
distributions (actually different ones, here distinguished by the argument
$A$ and $L$, respectively) independent of $T$. This scaling behavior represents
the known scaling of the mean values as function of walk length $T$.

\begin{figure}[htb]
 \includegraphics[width=0.45\textwidth]{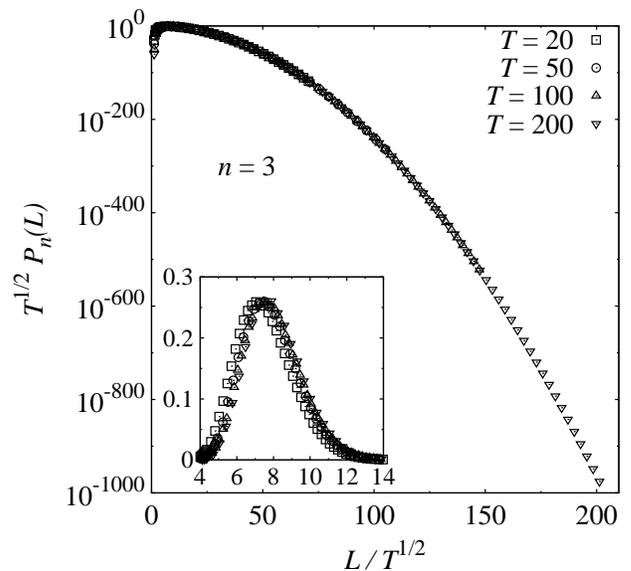}
\caption{Rescaled pdfs for $n=3$ walks, the perimeter $L$ of the convex hull 
 and different walk lengths $T$ in semi-logarithmic scale. Inset: Region 
 close to peaks in linear scale. \label{pdf_3walks_peri_res}}
\end{figure}

In Fig.\ \ref{pdf_3walks_area_res} the collapse according to 
Eq.~\eqref{scaling_A} is shown. In the tail of the rescaled pdfs almost 
perfect 
agreement of the curves for the different system sizes is visible. Only 
in the peak region (cf.\ inset of Fig.\ \ref{pdf_3walks_area_res}) small 
finite-size effects occur. The collapse of the pdfs for the perimeter in 
accordance with Eq.~\eqref{scaling_M} is depicted in 
Fig.~\ref{pdf_3walks_peri_res}. A good collapse with small finite-size 
deviations in the peak region (see inset) is also achieved.

Similar results were found and a good data collapse was achieved (not 
shown) for  the area and perimeter for $n=2$, respectively.

\subsection{Functional form of the probability density function}
According to \cite{Claussen2015} we use as universal distributions
$\widetilde{P}_n (\cdotp)$ in 
Eqs.~\eqref{scaling_A} and \eqref{scaling_M} two Gaussians
with mean $\mu_S$ and standard deviation $\sigma_S$ 
($S = L$ or $S = A$) in the case of large $T$. 
For the perimeter we obtain \cite{Claussen2015}:
\begin{equation}
 \widetilde{P}_n (m) = \frac{a_L}{\sqrt{2 \pi \, \sigma_L^2}} \; 
 \exp \left(- \frac{(m-\mu_L)^2}{2 \, \sigma_L^2} \right),
 \label{Gauss}
\end{equation}
where $a_L$ is a constant and $m = L/\sqrt{T}$. If we approximate 
$A \propto L^2$, so $l \propto m^2$ with $l = A/T$, we get an additional 
factor $1/\sqrt{l}$ from $|{\rm d} \, m / {\rm d} \, l| \sim 1/\sqrt{l}$ 
in the scaling relation. In total, the scaling function for the area is 
given by \cite{Claussen2015}:
\begin{equation}
 \widetilde{P}_n (l) = \frac{a_A}{\sqrt{2 \pi \, \sigma_A^2 \, l}} \; 
 \exp \left(- \frac{(\sqrt{l}-\mu_A)^2}{2 \, \sigma_A^2} \right),
 \label{Gauss_sqrt}
\end{equation}
where $a_A$ is some constant parameter.

\begin{figure}[htb]
 \includegraphics[width=0.45\textwidth]
 {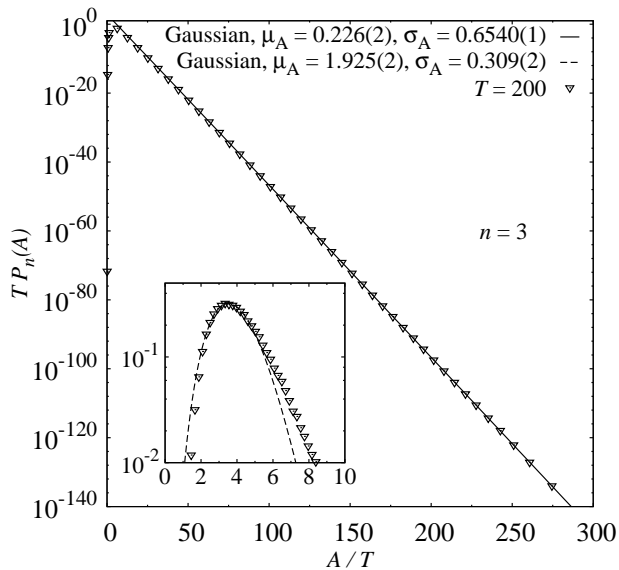}
\caption{Gaussian fit (solid line) according to Eq.~\eqref{Gauss_sqrt} 
with parameter $a_A=1000(25)$ to the rescaled pdf of the area for 
$T=200$ and $n=3$. 
Note that the 
logarithm of Eq.~\eqref{Gauss_sqrt} was fitted to the logarithm of the 
pdf for $A/T \geq 50$ to match the tail.
Inset: Gaussian fit (dashed line) in the peak region for $A/T \in [2,5]$ 
corresponding to Eq.~\eqref{Gauss_sqrt} with $a_A=0.474(3)$. 
\label{gauss_fit_area}}
\end{figure}

\begin{figure}[h!tb]
 \includegraphics[width=0.45\textwidth]
 {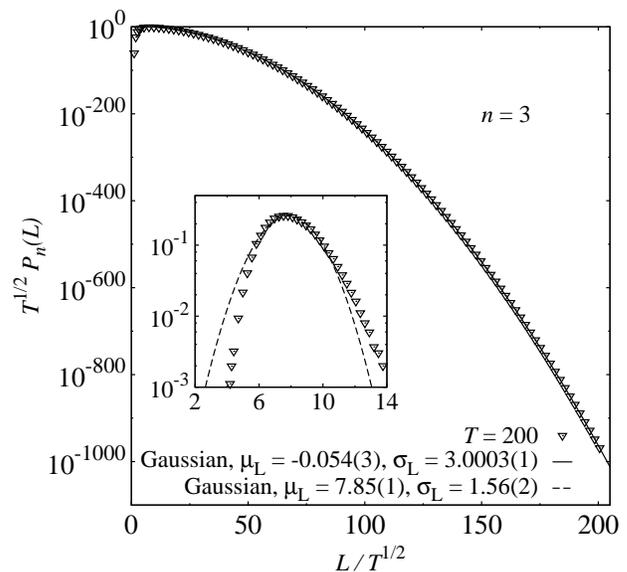}
\caption{Gaussian fit (solid line) according to Eq.~\eqref{Gauss} 
with parameter $a_L=1200(22)$ to the rescaled pdf of the perimeter 
for $T=200$ and $n=3$. 
Note that the logarithm of Eq.~\eqref{Gauss_sqrt} was fitted to the 
logarithm of the pdf for $L/\sqrt{T} \geq 20$ to match the tail.
Inset: Gaussian fit (dashed line) in the peak region for 
$L/\sqrt{T} \in [5,15]$ corresponding to Eq.~\eqref{Gauss} with 
$a_L=1.011(8)$. \label{gauss_fit_peri}}
\end{figure}

Figs.~\ref{gauss_fit_area} and \ref{gauss_fit_peri} show the results of 
those fits to the rescaled pdfs for $T=200$ and $n=3$. In both figures, 
two independent fits, one to the tail of the distribution and one to the 
peak region, were necessary. A single fit over the full support does not 
match the data. 
In Fig.~\ref{gauss_fit_area} both independent Gaussian fits fit the data 
well although deviations around the peak (cf., inset of 
Fig.~\ref{gauss_fit_area}) are visible.

Fig.~\ref{gauss_fit_peri}
shows the fits according to Eq.~\eqref{Gauss} to the rescaled pdf of the 
perimeter. The fit to the tail matches the region of large $L/\sqrt{T}$ 
very well. Nevertheless, in the inset of Fig.~\ref{gauss_fit_peri} strong 
deviations from a Gaussian behavior can be seen.

Next, we investigate the (left) tail of the pdfs towards small values of 
the rescaled area and perimeter. Corresponding to \cite{Claussen2015} we 
expect for the perimeter
asymptotically for small $m$ and small $T$ an essential singularity
according to
\begin{equation}
 \widetilde{P}_n(m) \sim a \, \exp \left( -\frac{b}{m^2} \right),
 \label{exp_peri}
\end{equation}
where $a$ and $b$ are constants and again $m = L/\sqrt{T}$. With similar 
arguments as for large $T$, where Gaussian fits are used (cf., 
Eqs.~\eqref{Gauss} and \eqref{Gauss_sqrt}) we obtain
for the small $l$ asymptotics of the area
\begin{equation}
 \widetilde{P}_n(l) \sim \frac{a}{\sqrt{l}} \, \exp \left( -\frac{b}{l} \right),
 \label{exp_area}
\end{equation}
where $a$ and $b$ are constants and again $l = A/T$. 

\begin{figure}[htb]
 \includegraphics[width=0.45\textwidth]
 {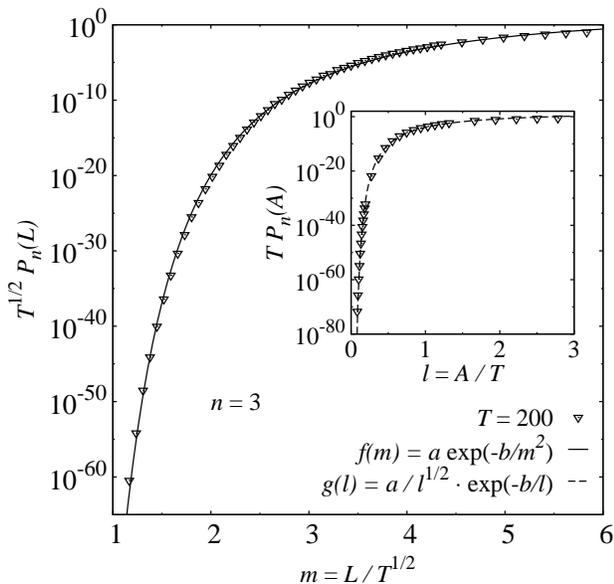}
\caption{Fit (solid line) with fit function $f(m)$, which 
corresponds to Eq.~\eqref{exp_peri} to 
the rescaled pdf of the perimeter for $T=200$ and $n=3$. Parameters of the 
fit are $a = 72(8)$ and $b = 200.9(5)$. Note that the logarithm of the  
fit function was fitted to the logarithm of the pdf for $m \in [1,5]$ to 
match the tail of the pdf. 
Inset: Fit (dashed line) with fit function $g(l)$
corresponding to Eq.~\eqref{exp_area} to 
the rescaled pdf of the area for $T=200$ and $n=3$. Parameters of the 
fit are $a = 394(22)$ and $b = 15.06(2)$. Note that the logarithm of the  
fit function was fitted to the logarithm of the pdf for $l \in [0,2]$ to 
match the tail of the pdf. \label{exp_fits}}
\end{figure}

Fig.~\ref{exp_fits} shows the fits to the left tails of the 
rescaled pdfs of the area and perimeter, respectively. The fit to the 
rescaled pdf of the perimeter matches the data quite well for small 
$m < 5$. Also the fit according to Eq.~\eqref{exp_area} (see inset of 
Fig.~\ref{exp_fits}) suits well to the rescaled pdf of the area for small 
$l < 2$.

\subsection{Rate function}
Next, the empirical rate function $\Phi_n (s)$ \cite{Touchette2009} is 
calculated which 
describes the leading behavior of the pdf in the large-deviation tail. If 
one assumes that the behavior of the probability density away from the 
typical values around the peak is exponentially small in the walk 
length~$T$, one gets for the rate function
\begin{equation}
 \Phi_n (s) = -\frac{1}{T} \ln P_n(s).
 \label{rate_fct}
\end{equation}
The quantity $s$ is usually normalized with the maximum possible values 
so that $s \in [0,1]$. As for Gaussian random walks no real maximum 
exists, we choose \cite{Claussen2015} $s_A = A/T^2$ and 
$s_L = L/T$, respectively.

\begin{figure}[htb]
 \includegraphics[width=0.44\textwidth]{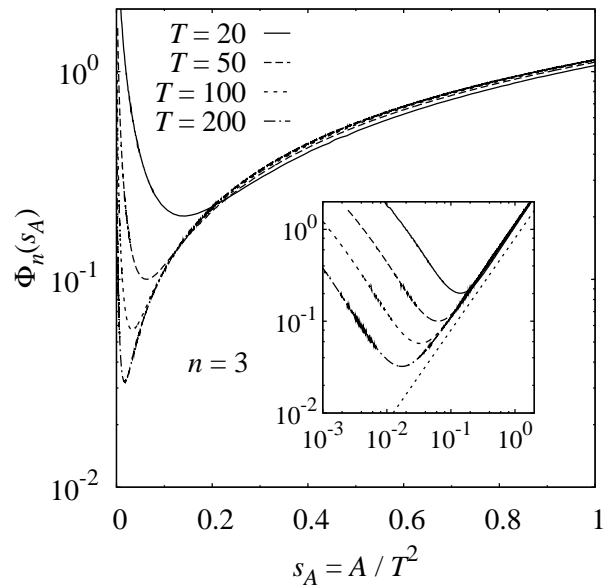}
\caption{Rate function $\Phi_n(s_A)$ as a function of the scaled area 
$s_A = A/T^2$ for different walk lengths $T$ and $n=3$ walks in 
semi-logarithmic scale. 
Inset: The same in a double-logarithmic plot, where the dashed line close 
to the data is a power law $s_A^{\kappa}$, with $\kappa = 1$. 
\label{rate_fct_3walks_area}}
\end{figure}

Fig.~\ref{rate_fct_3walks_area} shows the rate function for $n=3$ and the 
area of the convex hull. For small values of $s_A$ there are a strong 
finite length effects, whereas for larger values the curves for different 
$T$ seem to converge quickly to one curve. Nevertheless, a convergence
of the rate function to one universal shape seems likely, indicating that
the densities obey the large-deviation principle \cite{Touchette2009}.

To estimate  the behavior of the curves 
for large $T$ we plotted the rate function in a double-logarithmic scale 
in the inset of Fig.~\ref{rate_fct_3walks_area} and also show a power law 
$s_A^{\kappa}$ with $\kappa = 1$ for comparison. Apparently, our data has 
the same slope, at least in the region where $s_A$ is large. 
This is the same result as was found previously for $n=1$ \cite{Claussen2015}.

\begin{figure}[htb]
 \includegraphics[width=0.45\textwidth]{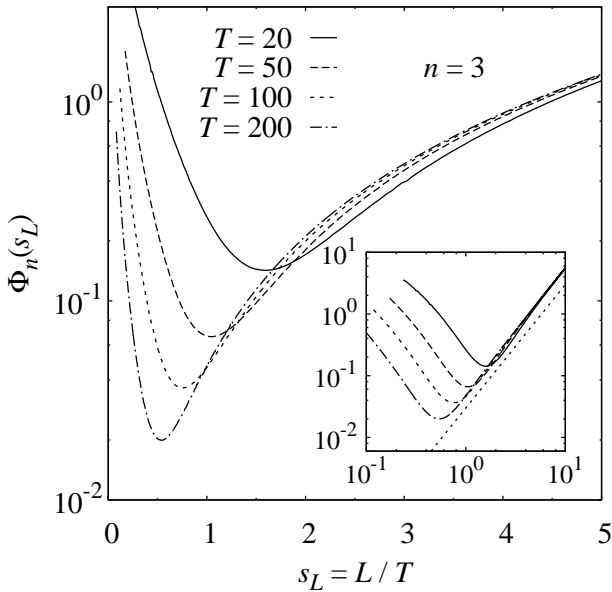}
\caption{Rate function $\Phi_n (s_L)$ as a function of the scaled perimeter 
$s_L = L/T$ for different walk lengths $T$ and $n=3$ walks in 
semi-logarithmic scale. 
Inset: The same in a double-logarithmic plot, where the dashed line close 
to the data is a power law $s_L^{\kappa}$, with $\kappa = 2$. 
\label{rate_fct_3walks_peri}}
\end{figure}

In Fig.~\ref{rate_fct_3walks_peri} the rate function for the perimeter 
and $n=3$ is depicted. Again, for small values of $s_L$ strong finite 
size effects occur and the convergence to a common curve is very slow. 
On the other hand, for larger $s_L$ the convergence to a common 
curve is already visible. In the inset of Fig.~\ref{rate_fct_3walks_peri} 
the data is shown in a double-logarithmic plot. Our data is compatible 
with a power-law behavior $s_L^{\kappa}$ with $\kappa=2$ for large 
$s_L$ represented by the dashed line.
Again we have found the same result as was found previously 
for $n=1$ \cite{Claussen2015}.

Thus, for few number of walks, 
the behavior of the rate functions for large $s$ of both the area and the 
perimeter agree with the expected ones \cite{Claussen2015} found for 
$n=1$. So, our pdfs are said to follow the ``large-deviation principle'' 
as they can be well described by a rate function given by 
Eq.~\eqref{rate_fct}. The behavior of the rate function when
increasing the number of walks more strongly is discussed below in the
following section.

\subsection{Scaling behavior with respect to the number $n$ of walks }

According to \cite{Majumdar2010} for $n$ two-dimensional open Brownian 
random walks the average area is expected to scale like
\begin{equation}
 \langle A_n \rangle  =  \beta_n T,
 \label{av_A}
\end{equation}
where the $n$-dependent prefactor is given by
\begin{equation}
 \beta_n = 4n \sqrt{\pi} \; \int_0^{\infty} u [\text{erf}(u)]^{n-1} \cdot [u e^{-u^2} - g(u)] \; \text{d}u,
 \label{beta_exact}
\end{equation}
$\text{erf}(u)$ is the error function
\begin{equation*}
 \text{erf}(u) = \frac{2}{\sqrt{\pi}} \int_0^u e^{-t^2} \; \text{d}t,
\end{equation*}
and
\begin{equation*}
 g(u) = \frac{1}{2 \sqrt{\pi}} \int_0^1 \frac{e^{-u^2/t}}{\sqrt{t (1-t)}} \; \text{d}t.
\end{equation*}
In the large-$n$ limit $\beta_n$ scales like \cite{Majumdar2010}
\begin{equation}
\beta_n \sim 2 \pi \, \ln n.
\label{eq:beta:n}
\end{equation}

According to \cite{Majumdar2010} the average perimeter of convex hulls of 
$n$ two-dimensional Brownian walks should scale like
\begin{equation}
 \langle L_n \rangle = \alpha_n \sqrt{T},
 \label{av_M}
\end{equation}
with
\begin{equation}
 \alpha_n = 4n \sqrt{2 \pi} \; \int_0^{\infty} u e^{-u^2} [\text{erf}(u)]^{n-1} \; \text{d}u,
 \label{alpha_exact}
\end{equation}
which has a large-$n$ scaling
\begin{equation}
\alpha_n \sim 2 \pi \, \sqrt{2\, \ln n}\,.
 \label{eq:alpha:n}
\end{equation}

\begin{figure}[ht]
 \includegraphics[width=0.45\textwidth]{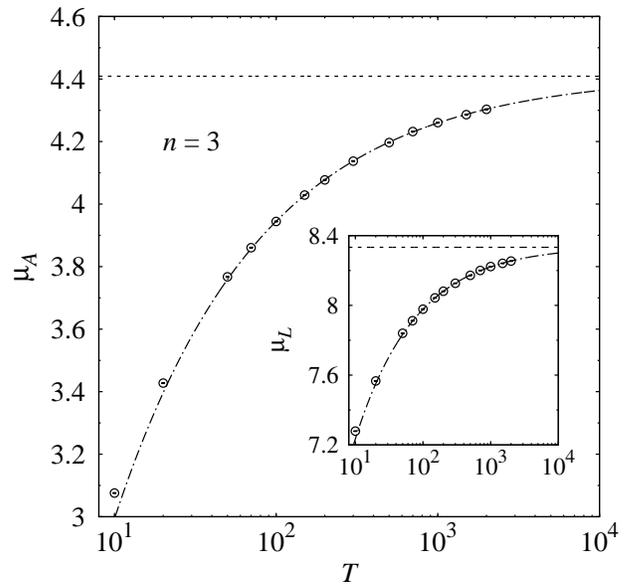}
\caption{Average rescaled area $\mu_A$ of the convex hull as a 
function of walk 
length $T$ for $n=3$ walks. Note the logarithmic scaling of the $T$-axis.
For each data point at least $8 \cdot 10^5$ 
samples were used. Dashed-dotted line is a power-law fit according to 
Eq.~\eqref{power_law}. Horizontal dashed line represents analytical 
expectation \cite{Majumdar2010} $\beta_3 \approx 4.410$. Inset: The same 
for the average rescaled perimeter $\mu_L$. Horizontal line denotes 
expectation according to \cite{Majumdar2010}: $\alpha_3 \approx 8.334$. 
\label{av_3walks}}
\end{figure}

To check these analytical predictions, we performed 
simple-sampling simulations to determine the average area and perimeter 
for various values  of $n\in[1,10^5]$. 
Fig.~\ref{av_3walks} shows the results for 
$n=3$ independent Gaussian walks and the rescaled averages 
$\mu_A = \langle A \rangle/T$ and 
$\mu_L = \langle L \rangle/\sqrt{T}$, 
where $\langle \cdotp \rangle$ denotes averaging. We simulated walk 
lengths $T \in [10,2000]$ and used at least $8 \cdot 10^5$ samples to 
determine the average. A power-law fit
\begin{equation}
 \mu_S(L) = \mu_S^{\infty} + a \cdot T^b
 \label{power_law}
\end{equation}
with parameter $\mu_S^{\infty}$ ($S=A$ or $S=L$) denoting the 
extrapolated value for $T \rightarrow \infty$, and fit parameters 
$a$, $b$ is performed. Excluding the small system sizes 
the fit is done over the range $T \in [150,2000]$, yielding a reduced 
chi-square value of $\chi^2_{\rm red} \approx 0.72$ for the area. One 
can observe a convergence towards the average area for infinite $T$, which 
is $\mu_A^{\infty} = 4.415(6)$. Compared to the 
literature~\cite{Majumdar2010} 
$\beta_3 = \pi + 3 - \sqrt 3 \approx 4.410$ the measured value agrees 
within error bars. The inset of Fig.~\ref{av_3walks} shows the fit to 
the rescaled average perimeter $\mu_L$. Again, a convergence 
towards the average for $T \rightarrow \infty$ is visible. The fit gives 
$\chi^2_{\rm red} \approx 0.76$ for $T \in [150,2000]$ with 
$\mu_L^{\infty} = 8.339(7)$. This value is compatible within error 
bars with the analytical derivation \cite{Majumdar2010} 
$\alpha_3 \approx 8.334$.

For $n=2$ walks we accomplished similar fits (not shown) as for $n=3$. 
The fit for the rescaled average area was performed with an reduced 
chi-square value of $\chi^2_{\rm red} \approx 0.20$ over system sizes 
$T \in [300,2000]$. The average value for an infinite system is 
$\mu_A^{\infty} = 3.144(5)$, which is compatible within error bars 
with literature \cite{Majumdar2010}: $\beta_2 = \pi \approx 3.142$. For 
the average perimeter we obtained by the fit according to 
Eq.~\eqref{power_law} 
$\chi^2_{\rm red} \approx 0.10$ with $\mu_L^{\infty} = 7.091(4)$. 
This value agrees within error bars with the published value 
\cite{Majumdar2010} $\alpha_2 = 4 \cdot \sqrt{\pi} \approx 7.090$.

\begin{figure}[htb]
 \includegraphics[width=0.45\textwidth]
 {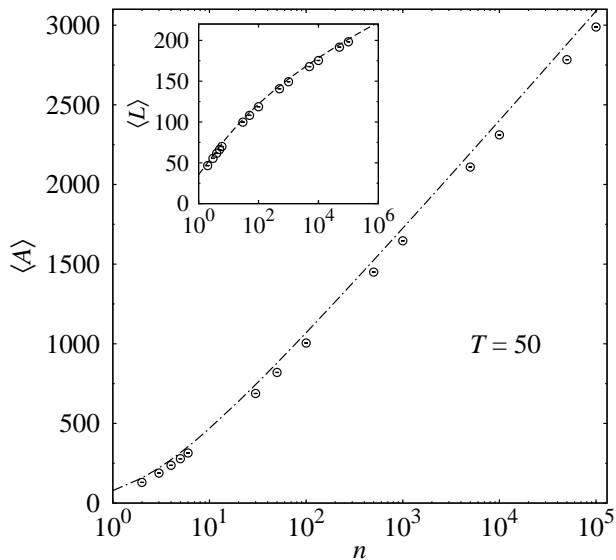}
\caption{Average area as a function of number of walks $n$ for 
$T=50$. Note the logarithmic scaling of the \mbox{$n$-axis}. For each data point 
about $10^6$ samples have been used. The dashed-dotted line 
shows the exact value given by Eqs.~\eqref{av_A} and \eqref{beta_exact}.
Inset: The same for the averaged perimeter. Dashed line 
displays exact value obtained by Eqs.~\eqref{av_M} and 
\eqref{alpha_exact}.
\label{av_L50_xwalks}}
\end{figure}

Next, we want to check if our data matches the exact equations 
\eqref{beta_exact} and \eqref{alpha_exact} for high values of $n$.
In Fig.~\ref{av_L50_xwalks} the averages of the area and perimeter 
obtained from simple-sampling simulations with $T=50$ fixed and various 
values for the number of walks $n$ is presented. 
Scaling the $n$-axis logarithmically 
leads to a linear behavior of the average area for large $n$ indicating a 
logarithmic dependence like expected by the previous scaling assumptions 
for $\beta_n$. As Eq.\ \eqref{beta_exact} is only valid for large values 
of $T$ one can see a small deviation between the theoretical and the 
measured values.

In the same way, the behavior of the average perimeter follows the
expected behavior, as shown in the inset of Fig.~\ref{av_L50_xwalks}.

Clearly, the data points for $T=50$ are located systematically
below the analytical curves, which is only valid for $T\to\infty$. 
This does not
come unexpectedly, because we see this behavior already for $n=3$ 
in Fig.~\ref{av_3walks}.
To check the convergence of the data for different walk lengths we 
investigate (Figures not shown) convex hulls for 
$n = 100$ walks as already done for $n=3$.
A fit according to Eq.~\eqref{power_law} for walk lengths 
$T \in [70,10^4]$ yields for the area a reduced chi-square value of 
$\chi^2_{\rm red} \approx 1.4$. One can observe a convergence for 
$T \rightarrow \infty$ towards $\mu_A^{\infty} = 21.40(1)$ 
which is 
compatible with the theoretical value (cf., Eq.~\eqref{beta_exact}) 
$\beta_{100} \approx 21.3890$ within a standard error bar. For the 
perimeter the power-law fit for \mbox{$T \in [100,10^4]$} results in 
$\chi^2_{\rm red} \approx 0.18$ 
and an extrapolated value for infinite $T$ which is 
$\mu_L^{\infty} = 17.262(1)$. Compared to the theoretical value 
$\alpha_{100} \approx 17.2596$ there is good agreement within two 
standard error bars. So, we can be confident that
our data shows the expected convergence for 
all values of $n$ towards the theoretical values.

\begin{figure}[htb]
 \includegraphics[width=0.45\textwidth]{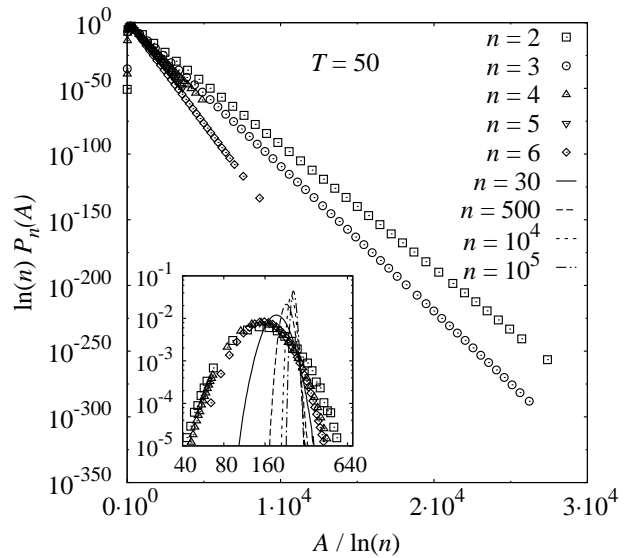}
\caption{Rescaled pdfs for $T=50$, the area $A$ of the convex 
 hull and various number of walks $n$ in semi-logarithmic scale. Inset: 
 Region close to peaks in double-logarithmic scale. Note that for walk 
 numbers $n \geq 30$ only values from simple sampling exist and therefore 
 only the region around the peak is depicted. 
 \label{pdf_L50_xwalks_area_res}}
\end{figure}

\begin{figure}[htb]
 \includegraphics[width=0.45\textwidth]{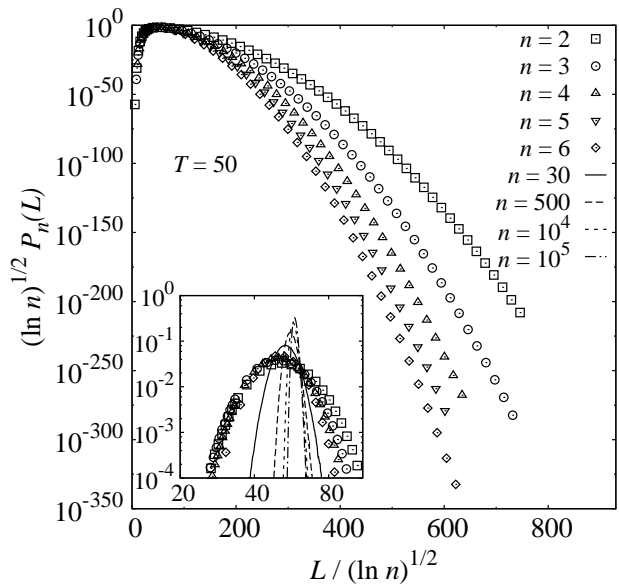}
\caption{Rescaled pdfs for $T=50$, the perimeter $L$ of the convex 
 hull and various number of walks $n$ in semi-logarithmic scale. Inset: 
 Region close to peaks in double-logarithmic scale. Note that for walk 
 numbers $n \geq 30$ only values from simple sampling exist and therefore 
 only the region around the peak is depicted. 
 \label{pdf_L50_xwalks_peri_res}}
\end{figure}

Next, we want to investigate whether the scaling behavior of the average 
with respect
to the number $n$ of walks transfers to the full distributions, as it
is the case with respect to the number $T$ of steps in the walks.
Figs.~\ref{pdf_L50_xwalks_area_res} and 
\ref{pdf_L50_xwalks_peri_res} show the distributions with a corresponding 
rescaling of the axis. Apparently the quality of the collapse is not very
good but seems to get gradually better when making the number $n$ of
walks very large. This can be seen when looking at the insets of
the figures, where the change of the distributions for $n=10^4$ $\to$
$n=10^5$ is rather small, compared to the change $n=2$ $\to$ $n=6$. This
corresponds to the just discussed behavior of the mean, where also
strong scaling corrections at small number $n$ of walks are visible. Thus,
a convergence to the scaling form, at rather large values of $n$, appears 
likely.

Finally, we consider the shape of the distributions in the limit
of a large number of walkers and long walk lengths. The distribution
of the perimeter, on which we focus here, 
can be approximated \cite{Majumdar2010} 
by the distribution of $2\pi$ times the
\emph{span} of $n$ independent one-dimensional random walkers.
Since the span is given by the sum of the two extreme points in positive and
negative directions  of the $n$ one-dimensional walkers, this distribution
is basically given by the convolution of two Gumbel extreme-value 
distributions. The
exact distribution \cite{kundu2013} for the one-dimensional case can
be formulated in terms of the modified Bessel function $K_0(x)$ of
zero'th order. Correspondingly to Eq.~(6) of Ref. \cite{kundu2013} we
fit our data for the perimeter to

\begin{equation}
f(L) = 2 a b \, \exp(-z) \, K_0 \, \left(2\exp(-z/2)\right)
\label{eq:Bessel:fit}
\end{equation}
with $b\equiv 2 \sqrt{\log n}$, $z\equiv b((L-B)/C-b)$. The variables  
$B$, $C$ and $a$ are fit parameters allowing for an adjustment of the
center and width of the distributions, and taking care of the normalization,
respectively.

\begin{figure}[htb]
 \includegraphics[width=0.45\textwidth]
 {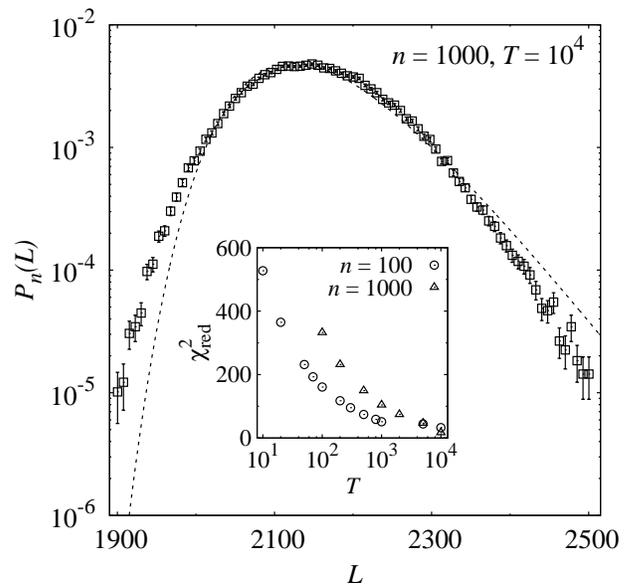}
\caption{Distribution of the perimeter for a larger number $n=1000$ of walkers
and a very long walk length $T=10^4$. The dashed line shows the result of a
fit to the distribution
given in Eq.~(\ref{eq:Bessel:fit}), wich resulted in  $a=3.83 \cdot 10^{-3}$, 
$B=754.96$, and $C=255.81$, yielding $\chi^2_{\rm red} = 16.84$.
The inset shows the reduced chi-square for two different numbers
of walkers as a function of the walk length.
 \label{bessel_fit}}
\end{figure}

In Fig.\ \ref{bessel_fit} the data for $n=1000$ and $T=10^4$ as obtained
from simple sampling is shown together with a corresponding fit. The fit
is good in the center of the distribution, but not away from it. Nevertheless,
as the inset shows,
when icreasing either $n$ or $T$ the quality of the fit increases 
considerably. 

\begin{figure}[htb]
 \includegraphics[width=0.45\textwidth]
 {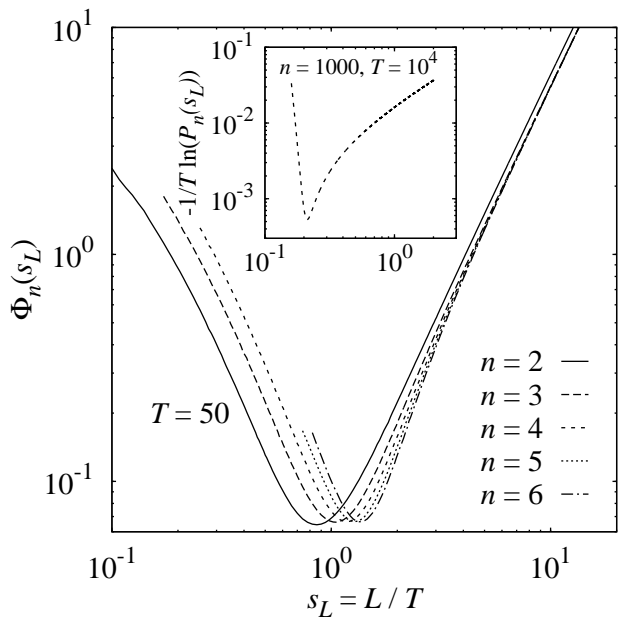}
\caption{Rate function $\Phi_n (s_L)$ as a function of the scaled 
perimeter $s_L = L/T$ for different number of walks $n \in [2,6]$ 
and walk length $T=50$ in double-logarithmic scale.
Inset: the function from Eq.\ (\ref{eq:Bessel:fit}) with the parameters
as obtained from the fit to the $n=1000$, $T=10^4$ data, rescaled as
to obtain a rate function, also shown in double-logarithmic scale.
\label{rate_fct_peri_xwalks}}
\end{figure}

Furthermore, we investiagted the  behavior of the rate function with varying 
number of walks $n$. We show $\Phi_n (s_L)$ in 
Fig.~\ref{rate_fct_peri_xwalks} as a function of $s_L = L/T$, where 
$T=50$ is fixed, for various values of $n$, while using here again
also the large-deviation
data.
For small $s_L$ a strong influence of $n$  can be 
seen, while this is weaker for larger values of $s_L$. 

In the inset of 
Fig.~\ref{rate_fct_peri_xwalks} we show the data of Eq.\ (\ref{eq:Bessel:fit})
with the parameters as obtained from the fit for $n=1000$ and $T=10^4$
plotted in the same way as the rate function, also in double-logarithmic
scale.
Apparently, the shape of the rate functions shown in the main plot 
become more and more
 similar to the function shown in the inset. The actual values are quite
different, because the values of $n$ and $T$ are very different for the two
cases. This is due to strong corrections to the leading scaling
behavior, as visible in 
Fig.\ \ref{rate_fct_3walks_area}, where also the minimum moves left and down
when increasing the walk length~$T$.
 Nevertheless, the result
 supports qualitatively the validity of Eq.\ (\ref{eq:Bessel:fit}).
 For a more quantitative statement
the numbers $n$ and $T$ which can be studied using the large-deviation approach
are too small due to the huge numerical effort which would needed in this case.

Anyway, our results indicate that for $n\to\infty$ and $T\to \infty$
the distribution of the perimeter of many walkers can indeed be
described by a suitably rescaled convolution of two Gumbel extreme-value
distributions.

For the case of the area (not shown) we observe a similar 
behavior as for the perimeter. 
Again we found that close to the minimum a change of the 
shape starts to appear. Nevertheless, here we have no functional
form available, so we do not discuss this further.

\section{Conclusion and Outlook \label{Concl}}

We have performed simulations of multiple two-dimensional 
discrete-time random walks
with Gaussian displacements. Convex hulls of the random walks have been 
calculated and the area $A$ and perimeter $L$ have been obtained. 
We have applied
a large-deviation scheme, via biasing Markov-chain Monte Carlo evolutions 
in the configuration space of walks. The bias was introduced with respect 
to large or small areas or perimeters, respectively. 
In this way we have been able to obtain these distributions, for
moderate number of walks, over large ranges of the support. Thus,
we could measure probability densities spanning as many as 1000
decades in probability. 
The resulting probability densities show the same scaling behavior as the mean
with respect to the length $T$ of the walks,
i.e., $\widetilde{P}_n (A/T)$ and $\widetilde{P}_n (L/\sqrt{T})$ appear to be 
universal densities.

For small numbers $n$ of walkers,
the shape of these universal densities follows Gaussian distributions 
for $L$ and $\sqrt{A}$, respectively, as for the $n=1$ case.
Also, for the deviations of the distribution in the direction of 
very small diameters and areas, the previously found ($n=1$) essential
singularity is obtained for low-$n$ multiple random walks.

We also obtained the rate functions
for area and perimeter, rescaled with the scaling behavior of the maxima, 
i.e., $T^2$ and $T$, respectively. 
For both quantities, the finite-length rate functions approach limiting
functions for $T\to\infty$, 
showing that the densities follow the large-deviation principle
\cite{denHollander2000,dembo2010}.
This makes it likely that using analytical approaches from 
large-deviation theory, some results for the distributions of the convex
hulls may be obtained.
Anyway, the rate functions seem to be well described by a
power law in the case $n\to \infty$, as found previously in the $n=1$ 
case.

Finally, we have verified, that the scaling behavior of the averages with
respect to the number $n$ of walks is predicted as in the literature
\cite{Majumdar2010}. The convergence is slow, such that
on the level of the full distribution the convergence to a limiting
function is not fully visible. Nevertheless, using simple-sampling
simulations of number of walks up to $n=10^5$, a convergence in the
high-probability, i.e., peak region is visible, making a full convergence
likely. Furthermore, these results are compatible with a convergence
of the distribution of the perimeter to a convolution of two Gumbel
extreme-value distributions.

For future research it would be interesting to investigate multiple 
interacting walkers~\cite{Acedo2002,Kundu2014}, or multiple walkers 
performing self-avoiding walks or loop-erased 
random walks~\cite{Lawler1987,Majumdar1992}. 
Furthermore, in higher dimensions a
change of the scaling of the obtained distributions and thus 
also of the shape of the 
distributions can be anticipated, making such studies useful. 
Finally, it would be 
very interesting to apply the methods used here to 
biological models to investigate the formation of animal 
territories~\cite{Giuggioli2011}, which had originally motivated this work.\\

\begin{acknowledgments}
This project was supported by the German Science Foundation (DPG)
under grant HA 3169/8-1.
The simulations were performed at the HERO cluster of the University
of Oldenburg funded by the DFG (INST 184/108-1 FUGG) and the 
ministry of Science and Culture (MWK) of the Lower Saxony State.
\end{acknowledgments}


\end{document}